\documentclass{ifacconf}
\usepackage[T1]{fontenc}
\usepackage[latin9]{inputenc}
\usepackage{amsmath}
\usepackage{amssymb}
\usepackage{graphicx}      % include this line if your document contains figures
\usepackage{natbib}        % required for bibliography

\makeatletter
\@ifundefined{showcaptionsetup}{}{%
\PassOptionsToPackage{caption=false}{subfig}}
\usepackage{subfig}
\makeatother
\begin{document}
\begin{frontmatter}

\title{A Level Set Approach to Online Sensing and Trajectory Optimization
with Time Delays\thanksref{footnoteinfo}} 

\thanks[footnoteinfo]{This research was supported by the Office of Naval Research under Grant N00014-18-WX01382.}

\author[First,Second]{Matthew R. Kirchner} 
%\author[Third]{Second B. Author, Jr.} 
%\author[Third]{Third C. Author}

\address[First]{ Image and Signal Processing Branch, Research Directorate, Naval Air Warfare Center Weapons Division, China Lake, CA 93555, USA (e-mail: matthew.kirchner@navy.mil)}
\address[Second]{Electrical and Computer Engineering Department, University of California, Santa Barbara, CA 93106, USA (e-mail: kirchner@ucsb.edu)}

\begin{abstract}                % Abstract of not more than 250 words.
Presented is a method to compute certain classes of Hamilton\textendash Jacobi
equations that result from optimal control and trajectory generation
problems with time delays. Many robotic control and trajectory problems
have limited information of the operating environment a priori and
must continually perform online trajectory optimization in real time
after collecting measurements. The sensing and optimization can induce
a significant time delay, and must be accounted for when computing
the trajectory. This paper utilizes the generalized Hopf formula,
which avoids the exponential dimensional scaling typical
of other numerical methods for computing solutions to the Hamilton\textendash Jacobi
equation. We present
as an example a robot that incrementally predicts a communication
channel from measurements as it travels. As part of this example,
we introduce a seemingly new generalization of a non-parametric formulation
of robotic communication channel estimation. New communication measurements
are used to improve the channel estimate and online trajectory optimization
with time-delay compensation is performed. 
\end{abstract}


\end{frontmatter}
%===============================================================================

\section{Introduction}
Time delays are common presence in real-world instantiations of dynamic
systems. This causes issues with stability and robustness when attempting
to design real-time control for these systems, and the
challenges of design and analysis of systems with time delays have
been well studied; see \cite{richard2003time}. Historically, there have
been many attempts to compensate for time delays in control theory,
such as the well-known Pad\'{e} approximation in classical linear
control theory \cite[Sec. 5.7.3]{franklin1994feedback}, which approximates
a pure delay as a rational transfer function or state transformations
such as presented in \cite{kwon1980feedback}. In a modern setting,
real-time optimal control (RTOC) [\cite{ross2006practical}] and model
predictive control (MPC) [\cite{camacho2013model}] have gained large
scale acceptance in control and trajectory optimization problems.
These seek to find a control sequence and the resulting trajectory that
minimizes a pre-defined cost functional. RTOC optimizes the cost functional
directly, while MPC computes on a finite time horizon and is frequently referred to as receding horizon control. Because
of this, MPC is sub-optimal and necessitates online re-computation
of the current state frequently, while RTOC only needs to be re-computed
online if the system is perturbed from the computed optimal trajectory.

Fast online computation is critical for these approaches, and delay
from computation becomes more apparent as the dimensionality and model
complexity of the optimization increases. In addition to time delays
induced from computation, many robotics problems have limited information
of the operating environment a priori. The robot must perform sensing
in real-time and as a result, necessitates online re-compuation of the optimal
trajectory with this new information. As methods for sensing in robotics
have become more sophisticated, the time delay induced becomes larger.
An illustrative example appeared in \cite{ali2016motion}, where a
complicated non-parametric model was used to estimate the quality
of a communication signal from measurements of
a transmitter with a known location. In that work, a RTOC scheme was
developed to minimize combined motion and communication energy of
the signal. The RTOC was re-computed every $10$ seconds and it was
noted that the combined sensing and trajectory optimization was around
$2$ seconds, or $20\%$ of the entire compute interval. A delay this
large can have drastic consequences if not accounted for. This was
not addressed in \cite{ali2016motion}.

As noted in \cite{lu2008delay}, attempts to directly account for
time delays in MPC formulations are limited, with the most notable
proposed in \cite{kwon2004general} for linear systems. However this
work is restricted to a specific LQR cost functional and does not
account for control saturation. We present in this paper a more general
approach based on Hamilton\textendash Jacobi theory, which provides
a natural way to deal with pure time delay in the control.

Generally, solutions to optimal control and trajectory problems can
be found by solving a Hamilton\textendash Jacobi (HJ) partial differential
equation (PDE) as it establishes a sufficient condition for optimality
[\cite{osmolovskii1998calculus}]. Traditionally, numerical solutions
to HJ equations require a dense, discrete grid of the solution space
as in \cite{osher2006level,mitchell2008flexible}. Computing
the elements of this grid scales poorly with dimension and has limited
use for problems with dimension greater than four. The exponential
dimensional scaling in optimization is sometimes referred to as the
``curse of dimensionality'' [\cite{bellman1957dynamic}].

A new result in \cite{darbon2016algorithms} discovered numerical
solutions based on the Hopf formula [\cite{hopf1965generalized}] that
do not require a grid and can be used to efficiently compute solutions
to a certain class of Hamilton\textendash Jacobi PDEs. However, that
only applied to systems with time-independent Hamiltonians of the
form $\dot{x}=f\left(u\left(t\right)\right)$, and has limited use
for general linear control problems. Recently, the classes of systems was expanded upon and generalizations
of the Hopf formula were used to solve optimal linear control problems
in high-dimensions  in \cite{kirchner2018primaldual} and differential
games as applied to a multi-vehicle, collaborative pursuit-evasion
problem in \cite{kirchner2017time}.

The HJ formulation of the trajectory problem allows a simple and direct
treatment of these computationally-induced time delays and there is
no need to resort to approximation schemes such as Pad\'{e}. The
main contribution of this paper is to generalize the Hopf formula
to directly account for time delays induced by online computation
of the optimal control and trajectory. Motivated by robotic vehicle
path planning problems where the communication is sensed and estimated
online, we present as an additional contribution a seemingly new non-parametric
model to estimate a communication channel where the location of the
transmitter is unknown a priori.

The rest of the paper is organized as follows. Section \ref{sec:Solutions-to-the HJ with Hopf}
reviews HJ theory as it relates to linear optimal control and presents
a level set method, based on the generalized Hopf formula, for fast
computation of optimal trajectories with known time delay. Section
\ref{sec:An-Example-of online sensing} presents an online trajectory
problem where a robot incrementally predicts a wireless communication
channel from measurements as it travels. As part of this example,
we derive a non-parametric model for channel estimation. Finally,
we present results from simulations of the method in Section \ref{sec:Results}.

\section{Solutions to the Hamilton\textendash Jacobi
Equation with the Hopf Formula}\label{sec:Solutions-to-the HJ with Hopf}

Before proceeding we introduce some notation and assumptions. We consider
linear dynamics
\begin{equation}
\frac{d}{ds}x\left(s\right)=Ax\left(s\right)+B\alpha\left(s\right),\,s\in\left[0,t\right],\label{eq:general dynamics}
\end{equation}
where $x\in\mathbb{R}^{n}$ is the system state and $\alpha\left(s\right)\in\mathcal{A}\subset\mathbb{R}^{m}$
is the control input, constrained to the convex admissible control
set $\mathcal{A}$. We let $\gamma\left(s;x,\alpha\left(\cdot\right)\right)\in\mathbb{R}^{n}$ denote a state trajectory that evolves in time, $s\in\left[0,t\right]$,
with control input sequence $\alpha\left(\cdot\right)\in\mathcal{A}$
according to $\left(\ref{eq:general dynamics}\right)$ starting from
initial state $x$ at $s=0$. The trajectory $\gamma$ is a solution
of $\left(\ref{eq:general dynamics}\right)$ in that it satisfies
$\left(\ref{eq:general dynamics}\right)$ almost everywhere:
\begin{align*}
\frac{d}{ds}\gamma\left(s;x,\alpha\left(\cdot\right)\right) & =A\gamma\left(s;x,\alpha\left(\cdot\right)\right)+B\alpha\left(\cdot\right),\\
\gamma\left(0;x,\alpha\left(\cdot\right)\right) & =x.
\end{align*}
We construct a cost functional for $\gamma\left(s;x,\alpha\left(\cdot\right)\right)$,
given terminal time $t$ as 
\begin{equation}
R\left(t,x,\alpha\left(\cdot\right)\right)=\int_{0}^{t}C\left(s,x,\alpha\left(s\right)\right)ds+J\left(\gamma\left(t;x,\alpha\left(\cdot\right)\right)\right),\label{eq: Cost Function}
\end{equation}
where the function $C:\left(0,+\infty\right)\times\mathbb{R}^{n}\times\mathbb{R}^{m}\rightarrow\mathbb{R}\cup\left\{ +\infty\right\} $
is the running cost and represents the rate that cost is accrued
over time. The value function $v:\mathbb{R}^{n}\times\left(0,+\infty\right)\rightarrow\mathbb{R}$
is defined as the minimum cost, $R$, among all admissible controls
for a given state $x$ with
\begin{equation}
v\left(x,t\right)=\underset{\alpha\left(\cdot\right)\in\mathcal{A}}{\text{inf}}\,R\left(t,x,\alpha\left(\cdot\right)\right).\label{eq: Value function}
\end{equation}
The value function in $\left(\ref{eq: Value function}\right)$ satisfies
the dynamic programming principle [\cite{bryson1975applied,evans10}]
and also satisfies the following initial value Hamilton\textendash Jacobi
(HJ) equation by defining the function $\varphi:\mathbb{R}^{n}\times\mathbb{R}\rightarrow\mathbb{R}$
as $\varphi\left(x,s\right)=v\left(x,t-s\right)$, with $\varphi$
being the viscosity solution of
\begin{equation}
\begin{cases}
\frac{\partial\varphi}{\partial s}\left(x,s\right)+H\left(s,x,\nabla_{x}\varphi\left(x,s\right)\right)=0,\\
\varphi\left(x,0\right)=J\left(x\right),
\end{cases}\label{eq:Initial value HJ PDE}
\end{equation}
where the Hamiltonian $H:\left(0,+\infty\right)\times\mathbb{R}^{n}\times\mathbb{R}^{n}\rightarrow\mathbb{R}\cup\left\{ +\infty\right\} $
is defined by
\begin{equation}
H\left(s,x,p\right)=\underset{\alpha\in\mathbb{R}^{m}}{\text{sup}}\left\{ \left\langle -f\left(s,x,\alpha\right),p\right\rangle -C\left(s,x,\alpha\right)\right\} .\label{eq: Basic Hamiltonian definition}
\end{equation}
The variable $p$ in\emph{ $\left(\ref{eq: Basic Hamiltonian definition}\right)$
}denotes the \emph{costate}, which in the HJ equation $\left(\ref{eq:Initial value HJ PDE}\right)$
is associated with the gradient of the value function. We denote by
$\lambda\left(s;x,\alpha\left(\cdot\right)\right)$ the costate trajectory
that satisfies almost everywhere: 
\begin{align*}
\frac{d}{ds}\lambda\left(s;x,\alpha\left(\cdot\right)\right) & =\nabla_{x}f\left(\gamma\left(s;x,\alpha\left(\cdot\right)\right),s\right)^{\top}\lambda\left(s;x,\alpha\left(\cdot\right)\right)\\
\lambda\left(t;x,\alpha\left(\cdot\right)\right) & =\nabla_{x}J\left(\gamma\left(t;x,\alpha\left(\cdot\right),s\right)\right),
\end{align*}
for $\forall s\in\left[0,t\right]$ with initial costate denoted by
$\lambda\left(0;x,\alpha\left(\cdot\right)\right)=p$. With a slight
abuse of notation, we will hereafter use $\lambda\left(s\right)$
to denote $\lambda\left(s;x,\alpha\left(\cdot\right)\right)$, since
the initial state and control sequence can be inferred through context
with the corresponding state trajectory, $\gamma\left(s;x,\alpha\left(\cdot\right)\right)$.

\subsection{Viscosity Solutions with the Hopf Formula}

Consider simplified system dynamics represented as

\begin{equation}
\frac{d}{ds}x\left(s\right)=f\left(\alpha\left(s\right)\right).\label{eq: Basic diff eq}
\end{equation}
The associated HJ equation no longer depends on state and is given
as
\begin{equation}
\begin{cases}
\frac{\partial\varphi}{\partial s}\left(x,s\right)+H\left(\nabla_{x}\varphi\left(x,s\right)\right)=0,\\
\varphi\left(x,0\right)=J\left(x\right).
\end{cases}\label{eq:Non-state dependent HJ}
\end{equation}
When $J\left(x\right)$ is convex and continuous in $x$, and $H\left(p\right)$
is continuous in $p$, it was shown in \cite{evans10} that an exact,
point-wise viscosity solution to $\left(\ref{eq:Non-state dependent HJ}\right)$
can be found using the Hopf formula [\cite{hopf1965generalized}]
\begin{equation}
\varphi\left(x,t\right)=-\underset{p\in\mathbb{R}^{n}}{\text{min}}\left\{ J^{\star}\left(p\right)+tH\left(p\right)-\left\langle x,p\right\rangle \right\} ,\label{eq: Basic Hopf formula}
\end{equation}
with the Fenchel\textendash Legendre transform, denoted $J^{\star}:\mathbb{R}^{n}\rightarrow\mathbb{R}\cup\left\{ +\infty\right\} $,
defined for a convex, proper, lower semicontinuous function $J:\mathbb{R}^{n}\rightarrow\mathbb{R}\cup\left\{ +\infty\right\} $
[\cite{hiriart2012fundamentals}] as
\begin{equation}
J^{\star}\left(p\right)=\underset{x\in\mathbb{R}^{n}}{\text{sup}}\left\{ \left\langle p,x\right\rangle -J\left(x\right)\right\} .\label{eq: Fenchel transform}
\end{equation}
The transform defined in $\left(\ref{eq: Fenchel transform}\right)$
is also referred to in literature as the \emph{convex conjugate}.

Proceeding similar to \cite{kirchner2017time}, we can generalize
the Hopf formula to $\left(\ref{eq:general dynamics}\right)$
by making a change of variables
\begin{equation}
z\left(s\right)=e^{-sA}x\left(s\right),\label{eq:change of varibles}
\end{equation}
which results in the following system
\begin{equation}
\frac{d}{ds}z\left(s\right)=f\left(s,\alpha\left(s\right)\right)=e^{-sA}B\alpha\left(s\right).\label{eq:z transformed system}
\end{equation}
The terminal cost function is now defined in $z$ with
\begin{equation}
\varphi\left(z,0\right)=J\left(e^{tA}z\right).\label{eq:Terminal cost as function of z}
\end{equation}
For clarity in the sections to follow, we use the notation $\widehat{H}$
to refer to the Hamiltonian for systems defined by $\left(\ref{eq:z transformed system}\right)$
and $H$ for systems defined by $\left(\ref{eq:general linear system}\right)$.
Notice that the system $\left(\ref{eq:z transformed system}\right)$
does not depend on state but is now time-varying. It was shown in
\cite[Section 5.3.2, p. 215]{kurzhanski2014dynamics} that the Hopf
formula can be generalized for a time-dependent Hamiltonian of the system in $\left(\ref{eq:z transformed system}\right)$
with
\begin{align}
\varphi\left(x,t\right) & =-\underset{p\in\mathbb{R}^{n}}{\text{min}}\Bigg\{ J^{\star}\left(e^{-tA^{\top}}p\right)\label{eq:generalized hopf formula}\\
 & +\int_{0}^{t}\widehat{H}\left(s,p\right)ds-\left\langle x,p\right\rangle \Bigg\},\nonumber 
\end{align}
with Hamiltonian defined by
\[
\widehat{H}\left(s,p\right)=\underset{\alpha\in\mathbb{R}^{m}}{\text{sup}}\left\{ \left\langle -e^{-sA}B\alpha,p\right\rangle -C\left(s,\alpha\right)\right\} .
\]
For the remainder of the paper, we consider a time-optimal formulation
in which $\widehat{H}$ is defined as
\begin{equation}
\widehat{H}\left(s,p\right)=\underset{\alpha\in\mathbb{R}^{m}}{\text{sup}}\left\{ \left\langle -e^{-sA}B\alpha,p\right\rangle -\mathcal{I}_{\mathcal{A}}\left(\alpha\right)\right\} ,\label{eq:transformed hamiltonian with indicator cost}
\end{equation}
where $\mathcal{I}_{\mathcal{A}}:\mathbb{R}^{m}\rightarrow\mathbb{R}\cup\left\{ +\infty\right\} $
is the indicator function for the set $\mathcal{A}$ and is defined
by
\[
\mathcal{I}_{\mathcal{A}}\left(\alpha\right)=\begin{cases}
0 & \text{if}\,\alpha\in\mathcal{A}\\
+\infty & \text{otherwise.}
\end{cases}
\]
Suppose $\mathcal{A}$ is a closed convex set such that $0\in\text{int}\,\mathcal{A}$,
where $\text{int}\,\mathcal{A}$ denotes the interior of the set $\mathcal{A}$.
Then $\left(\mathcal{I}_{\mathcal{A}}\right)^{\star}$ defines a norm,
which we denote with $\left\Vert \left(\cdot\right)\right\Vert _{\mathcal{A}^{*}}$,
which is the dual norm [\cite{hiriart2012fundamentals}] to $\left\Vert \left(\cdot\right)\right\Vert _{\mathcal{A}}$.
From this we can write $\left(\ref{eq:transformed hamiltonian with indicator cost}\right)$
in general as
\begin{equation}
\widehat{H}\left(s,p\right)=\left\Vert -B^{\top}e^{-sA^{\top}}p\right\Vert _{\mathcal{A}^{*}}.\label{eq: Hamiltonian dual norm in z}
\end{equation}

\subsection{Time-Optimal Control to a Goal Set\label{subsec:Time-Optimal-Control-to}}

Consider a goal set $\Omega\subset\mathbb{R}^{n}$ and a task to determine
the control that drives the system into $\Omega$ in minimal time.
We represent the set $\Omega$ as an implicit surface with cost function
$J:\mathbb{R}^{n}\rightarrow\mathbb{R}$ such that
\[
\begin{cases}
J\left(x\right)<0 & \text{for any}\,x\in\text{int}\,\Omega,\\
J\left(x\right)>0 & \text{for any}\,x\in\left(\mathbb{R}^{n}\setminus\Omega\right),\\
J\left(x\right)=0 & \text{for any}\,x\in\left(\Omega\setminus\text{int}\,\Omega\right),
\end{cases}
\]
where $\text{int}\,\Omega$ denotes the interior of $\Omega$. Note
that if the goal is a point in $\mathbb{R}^{n}$, then we can represent
this by choosing $\Omega$ as a ball with arbitrarily small radius.
As noted in \cite{kirchner2018primaldual} we solve for the minimum
time to reach the set $\Omega$ by constructing a newton iteration,
starting from an initial guess, $t_{0}$, with
\begin{equation}
t_{i+1}=t_{i}-\frac{\varphi\left(x,t_{i}\right)}{\frac{\partial\varphi}{\partial t}\left(x,t_{i}\right)},\label{eq:Newton iterate}
\end{equation}
where $\varphi\left(x,t_{i}\right)$ is the solution to $\left(\ref{eq:generalized hopf formula}\right)$
at time $t_{i}$. Notice the value function mush satisfy the HJ equation
\[
\frac{\partial\varphi}{\partial t}\left(x,t_{i}\right)=-H\left(\nabla_{x}\varphi\left(x,t_{i}\right),x\right),
\]
where $\nabla_{x}\varphi\left(x,t_{i}\right)$ is the argument of
the minimizer in $\left(\ref{eq:generalized hopf formula}\right)$,
which we will denote as $p^{*}$. No change of variable is needed
for the Newton update and we have
\[
H\left(p^{*},x\right)=-x^{\top}A^{\top}p^{*}+\left\Vert -B^{\top}p^{*}\right\Vert _{*}.
\]
We iterate $\left(\ref{eq:Newton iterate}\right)$ until convergence
at the optimal time to reach, which we denote as $t^{*}$. The optimal
control can be found directly from the necessary conditions of optimality
established by Pontryagin's principal [\cite{pontryagin2018mathematical}]
by noting the optimal trajectory, denoted as $\gamma^{*}$, must satisfy
\begin{align*}
\frac{d}{ds}\gamma^{*}\left(s;x,\alpha^{*}\left(\cdot\right)\right) & =-\nabla_{p}H\left(\lambda^{*}\left(s\right)\right)\\
 & =A\gamma\left(s;x,\alpha^{*}\left(\cdot\right)\right)\\
 & +B\nabla_{p}\left\Vert -B^{\top}\lambda^{*}\left(s\right)\right\Vert _{*},
\end{align*}
where $\lambda^{*}$ is the optimal costate trajectory and is given
by
\[
\lambda^{*}\left(s\right)=e^{-sA^{\top}}p^{*}.
\]
This implies our optimal control is
\[
\alpha^{*}\left(s\right)=\nabla_{p}\left\Vert -B^{\top}e^{-sA^{\top}}p^{*}\right\Vert _{*},
\]
for all time $s\in\left[0,t^{*}\right]$, provided the gradient exists.

Note that in the above formulation, we compute a viscosity solution
to $\left(\ref{eq:Non-state dependent HJ}\right)$ without constructing
a discrete grid and this method can provide a numerical solution that
is efficient to compute even when the state space is high-dimensional.
Additionally, no derivative approximations are needed with Hopf formula-based
methods, and this eliminates the numeric dissipation introduced with
the Lax\textendash Friedrichs scheme that is necessary to maintain
stability in grid-based methods.

\section{Online Trajectory Optimization With Delay}\label{subsec:General-Linear-Models}

%\begin{figure*}
%\begin{centering}
%\subfloat[The level set evolution of the double integrator system with a single
%$0.1\,sec$ time delay.]{\centering{}\includegraphics[height=6cm]{LevelSet0_1Sec}}~\subfloat[Level set evolution of the same system with a single $0.5\,sec$ time
%delay.]{\begin{centering}
%\includegraphics[height=6cm]{LevelSet0_5Sec}
%\par\end{centering}
%}
%\par\end{centering}
%\caption{Level set evolution of a double integrator system without time delay
%is shown in black. The level set computed of the double integrator
%system with time delay is shown in red. Both were computed at 10 equally
%spaced times on the interval from zero to two seconds.\label{fig:Level set with time delay.}}
%\end{figure*}

We'll be considering a RTOC framework for the rest of this work with
a variable time horizon $s\in\left[0,t^{k}\right]$, where each $t^{k}$
is the optimal time-to-go as computed by Section \ref{subsec:Time-Optimal-Control-to}.
This can easily be applied to MPC problems by using a fixed, finite
time horizon $s\in\left[0,t\right]$. The superscript $k\in\mathbb{N}$
is used to denote the computational update of each relevant quantity,
with $k=0$ being the first optimization. With this notation, $x^{k}$
denotes our initial state when we initiate the computation, and likewise
we denote by $\gamma^{*}\left(s;x^{k},\alpha^{k}\left(\cdot\right)\right)$
as the $k$-th computed optimal trajectory. We re-compute the control
online after $\delta^{k}$ seconds of traveling on the trajectory
$\gamma^{*}\left(s;x^{k},\alpha^{k}\left(\cdot\right)\right)$, which
gives the new initial state for the next optimization as
\[
x^{k+1}=\gamma^{*}\left(\delta^{k};x^{k},\alpha^{k}\left(\cdot\right)\right).
\]
Using $x^{k+1}$ as the new initial condition, we proceed to use the
methods of Section \ref{subsec:Time-Optimal-Control-to} to compute
$\alpha^{k+1}\left(s\right)$. Note that the time between updates,
$\delta^{k}$ need not be uniform. Now consider the following linear
state space model with time delay
\begin{equation}
\frac{d}{ds}x\left(s\right)=Ax\left(s\right)+Bu^{k}\left(s-\tau^{k}\right),\label{eq:general linear system}
\end{equation}
with delayed control input $u^{k}\in\mathbb{\mathcal{A}\subset R}^{m}$,
and $\tau^{k}$ is the step-specific time delay. We can represent
the delay dynamics in the same form as $\left(\ref{eq:general dynamics}\right)$
by defining $\alpha^{k}\left(\cdot\right)$, for $k\neq0$, as
\[
\alpha^{k}\left(s\right)=\begin{cases}
u^{k}\left(s-\tau^{k}\right) & s\geq\tau\\
\alpha^{k-1}\left(s+\delta^{k-1}\right) & s<\tau.
\end{cases}
\]
We only consider causal systems and therefore assume each $\tau^{k}\geq0$
and
\[
\alpha^{0}\left(s\right)=\begin{cases}
u^{0}\left(s-\tau^{0}\right) & s\geq\tau\\
0 & s<\tau.
\end{cases}
\]
The Hamiltonian becomes
\begin{equation}
\widehat{H}\left(s,p\right)=\begin{cases}
\left\Vert -B^{\top}e^{-sA^{\top}}p\right\Vert _{\mathcal{A}^{*}} & s\geq\tau\\
-p^{\top}e^{-sA}B\alpha^{k-1}\left(s+\delta^{k-1}\right) & s<\tau
\end{cases},\label{eq:Hamiltonian of 'z'}
\end{equation}
for the Hopf formula in $\left(\ref{eq:generalized hopf formula}\right),$
with $p^{k}$ denoting the optimal inital costate for the update.
Likewise we solve for the optimal time-to-reach of the delayed system
using the methods of Section $\left(\ref{subsec:Time-Optimal-Control-to}\right)$.
The Hamiltonian for the Newton iteration becomes
\begin{equation}
H\left(p^{k},x\right)=\begin{cases}
-x^{\top}A^{\top}p^{k}+\left\Vert -B^{\top}p^{k}\right\Vert _{\mathcal{A}^{*}} & s\geq\tau\\
-x^{\top}A^{\top}p^{k}-\left(p^{k}\right)^{\top}B\alpha^{k-1}\left(s+\delta^{k-1}\right) & s<\tau
\end{cases}.\label{eq:Hamiltonian for Newton and delay}
\end{equation}
This implies our control becomes
\[
\alpha^{k}\left(s\right)=\begin{cases}
\nabla_{p}\left\Vert -B^{\top}e^{-sA^{\top}}p^{k}\right\Vert _{*} & s\geq\tau\\
\alpha^{k-1}\left(s+\delta^{k-1}\right) & s<\tau.
\end{cases}
\]

%Figure \ref{fig:Level set with time delay.} shows the zero level
%set of the value function at different times for a double integrator
%system with and without delay, illustrating the extreme effect
%time delays can have on control optimization problems.

\section{An Example of Online Sensing
and Trajectory Optimization}\label{sec:An-Example-of online sensing}

\begin{figure*}
\begin{centering}
\subfloat[Channel estimate with 5 measurements along shown path.]{\centering{}\includegraphics[width=5.5cm]{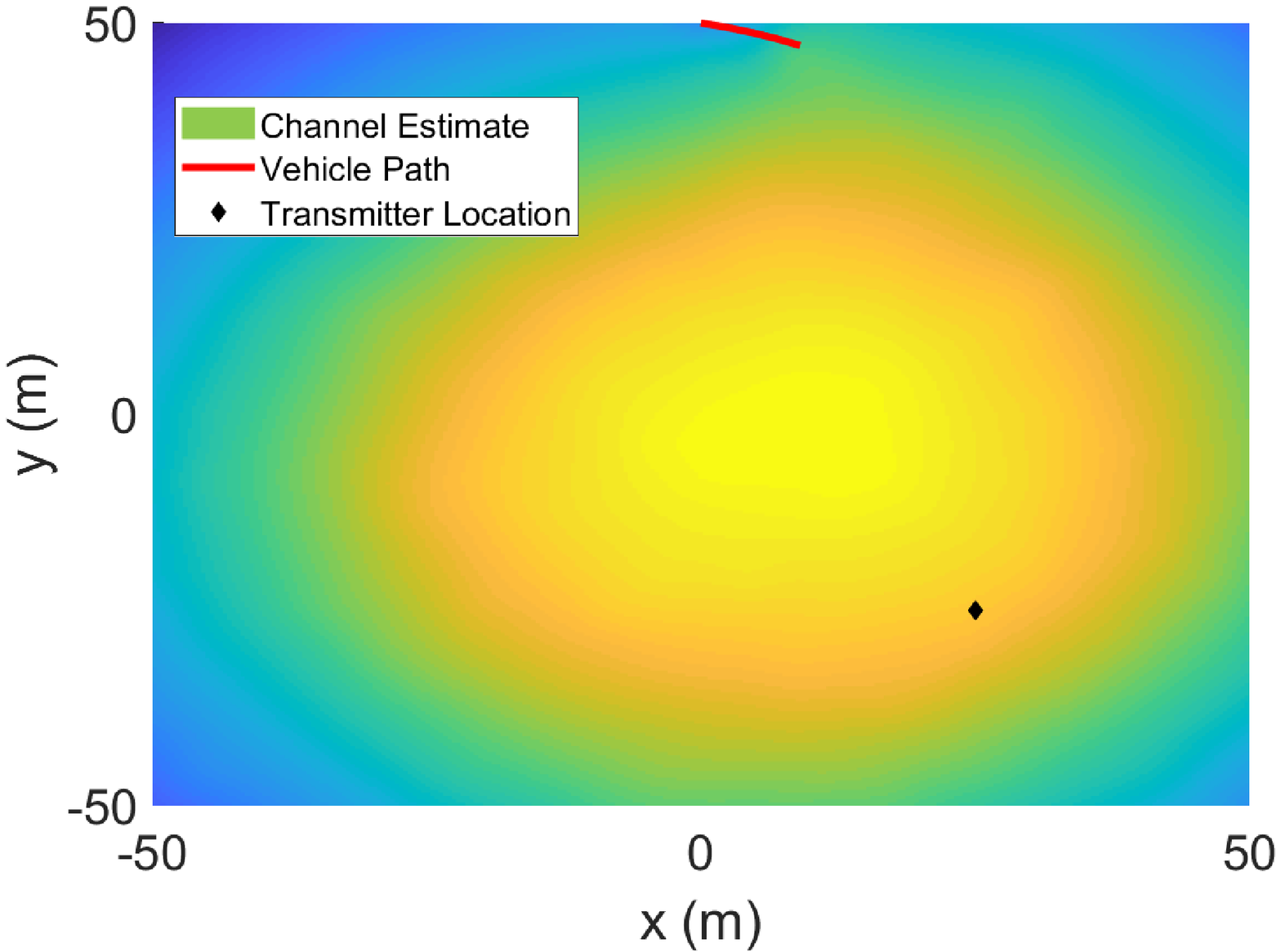}}\hspace*{\fill}\subfloat[Channel estimate with 30 measurements along shown path.]{\begin{centering}
\includegraphics[width=5.5cm]{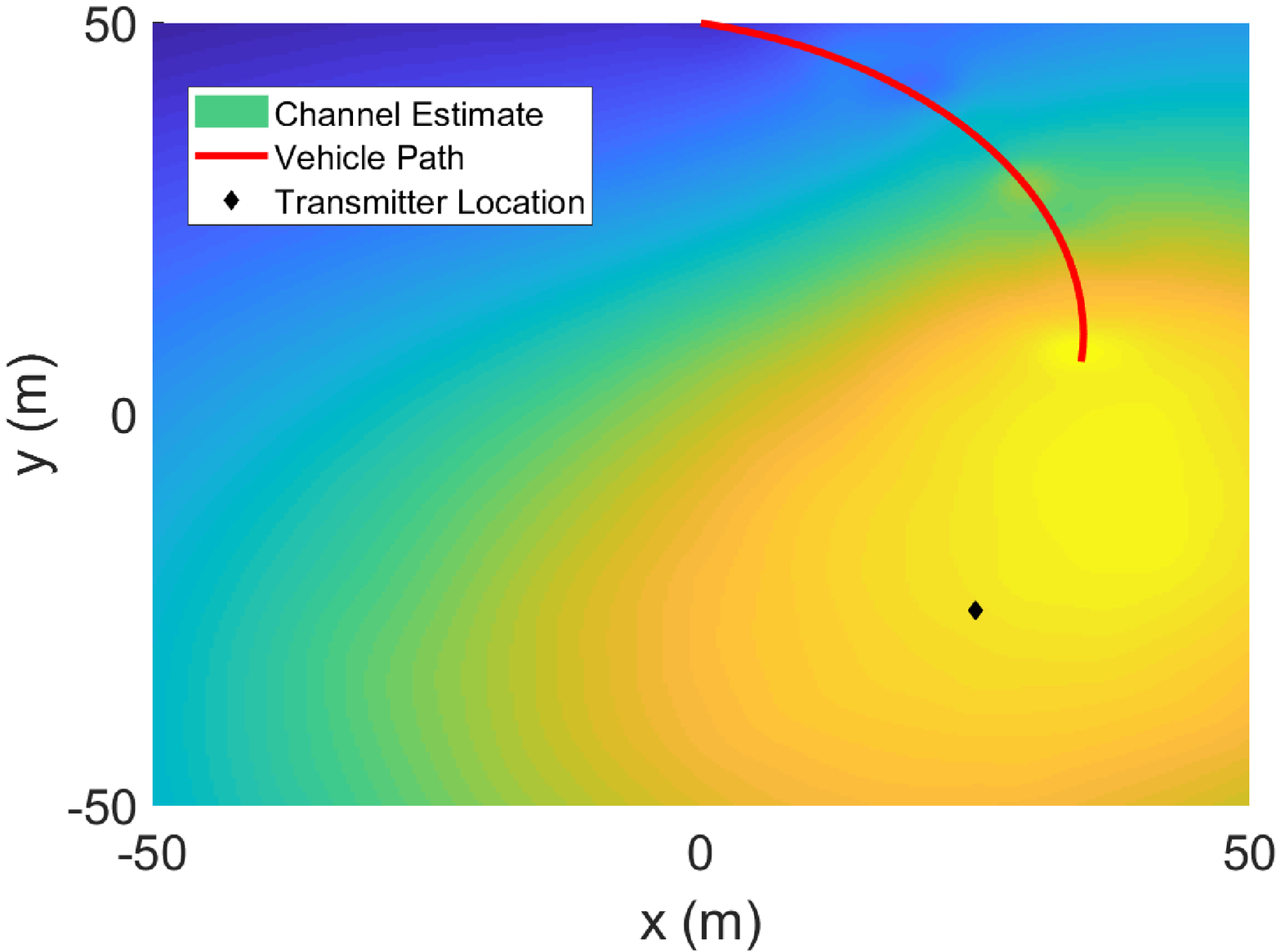}
\par\end{centering}
}\hspace*{\fill}\subfloat[Channel estimate with 75 measurements along shown path.]{\centering{}\includegraphics[width=5.5cm]{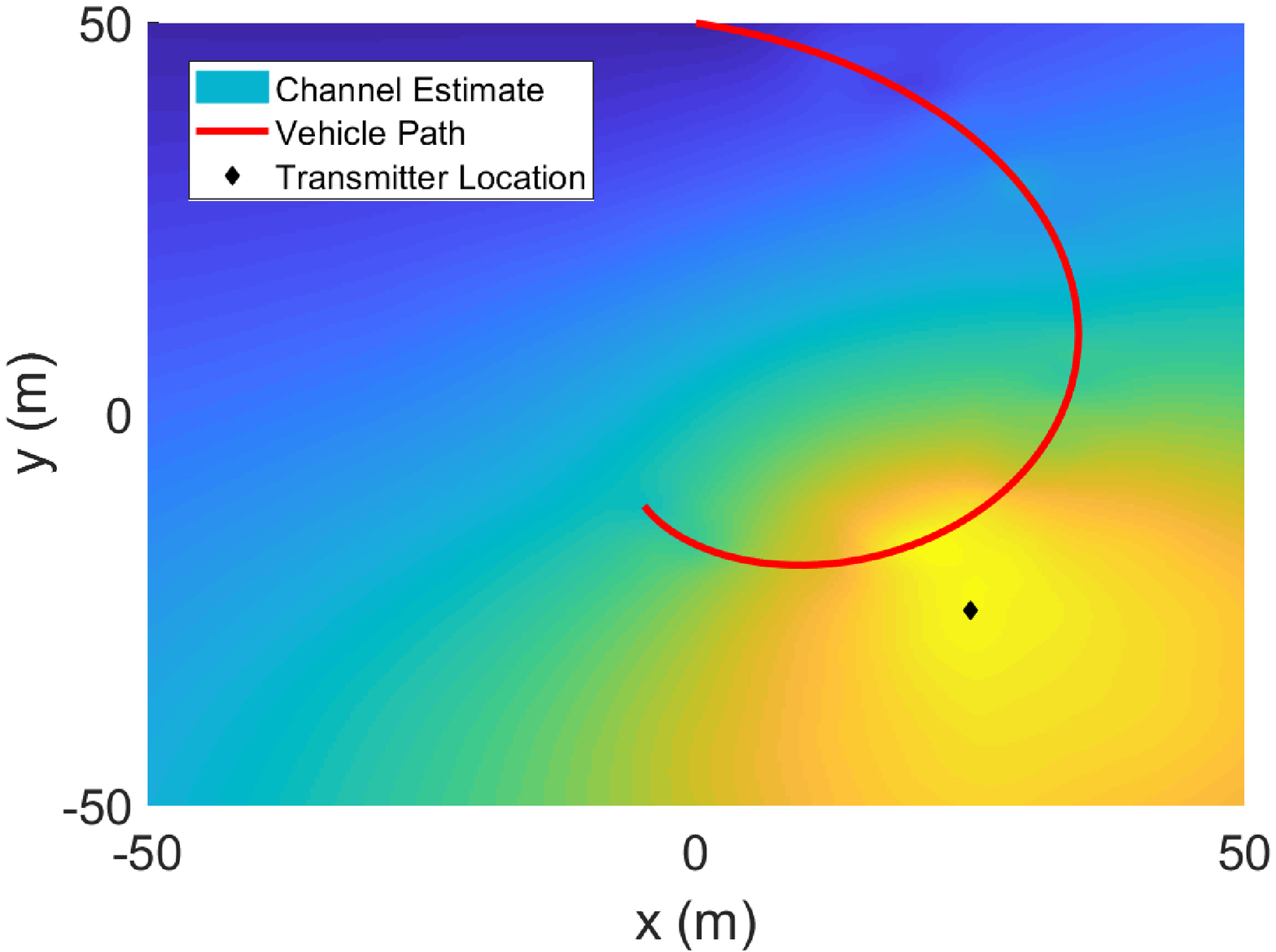}}
\par\end{centering}
\caption{The channel signal estimates from measurements as a vehicle traverses a known
path. The vehicle path is shown in red and the ground-truth transmitter
location shown as a black diamond. Best viewed in color. \label{fig:Channel est example}}
\end{figure*}

We present as an example a robotic vehicle that uses a radio transmitter
to communicate data to a remote base station. The goal is to plan
a trajectory to deliver the robot to a location with the best possible
communication performance, in minimum time. The location of base station
is unknown a priori and hence the communication link performance needs
to be spatially estimated. Initially, the robot will have only a sparse
number of samples of the communication link, or perhaps none at all,
and needs to determine the channel-to-noise ratio (CNR). As the robot
moves, more samples are collected and the estimate is improved. Each
time the estimate is improved, a new trajectory needs to be generated,
since the original trajectory is no longer optimal under the previous
estimate.

This is similar to the problem addressed in \cite{ali2016motion},
but that work attempted to minimize total power, was not time-optimal, and did not account for time delays.
We choose this example to demonstrate the method presented since
communication link performance needs to be sensed online and the computation time for the channel estimation is non-trival.
In the case of \cite{ali2016motion}, 2 seconds of computation was
required for every 10 second computation cycle.

\subsection{Channel Estimation with Unknown Emitter Location}

It was proposed in \cite{malmirchegini2010spatial} to model the CNR
(in dB) at a particular spatial location $q\in\mathbb{R}^{2}$ from
$\ell$ observed measurements as
\begin{equation}
\Upsilon_{dB}\left(q\right)=\Gamma\left(q;q_{b}\right)+\Delta\left(q\right),\label{eq:basic CNR model}
\end{equation}
where $\Gamma\left(q;q_{b}\right)$ is a parametric model of path
loss, relating CNR to spatial distance to the (known) transmitter
location, $q_{b}$ and is given as
\[
\Gamma\left(q;q_{b}\right)=c_{PL}-10n_{PL}\log\left(\left\Vert q-q_{b}\right\Vert \right),
\]
where $c_{PL}$ and $n_{PL}$ are constant parameters associated with the transmitter. The quantity $\Delta\left(q\right)$ represents
the deviation of the true CNR from $\Gamma$ and is modeled nonparametrically
as a Gaussian process [\cite{rasmussen2004gaussian}] with
\[
\Delta\left(q_{i}\right)\sim\mathcal{GP}\left(0,k_{\Delta}\left(q_{i},q_{i}\right)\right),
\]
for each $q_{i}$, where $k_{\Delta}$ is referred to as the \emph{covariance
function} and for modeling communication channels was suggested in
\cite{malmirchegini2010spatial} to be 
\[
k_{\Delta}\left(q_{i},q_{j}\right)=\xi^{2}e^{-\frac{\left\Vert q_{i}-q_{j}\right\Vert }{\eta}}+\sigma_{\rho}^{2},
\]
for any $i,j\in\left\{ 1,\ldots,\ell\right\} $ channel measurements.
The idea of constructing a model as a composition of a known parametric
model and a non-parametric deviation is not uncommon and is generally
formulated as Gaussian process regression with explicit basis functions
\cite[Chapter 2.7]{rasmussen2004gaussian}. We present the model while
omitting a detailed derivation as this is outside the scope of this
work. For more information, the reader is encouraged to review \cite{rasmussen2004gaussian}
for a thorough and complete review of Gaussian process regression
and \cite{malmirchegini2010spatial} for the derivation of the proposed
covariance functions for communication models. 

The model $\left(\ref{eq:basic CNR model}\right)$ assumes a priori
knowledge of the location of the transmitter, which for many robotic
path planning problems may not be available at run-time. We propose
to generalize $\left(\ref{eq:basic CNR model}\right)$ by considering
the case where the transmitting location is an unknown random variable
distributed with probability density function $g\left(q_{b}\right)$
and model $\Upsilon$ with a single Gaussian process with
\[
\Upsilon_{dB}\left(q\right)\sim\mathcal{GP}\left(0,k\left(q,q_{j}\right)\right),
\]
where
\[
k\left(q_{i},q_{j}\right)=k_{\Delta}\left(q_{i},q_{j}\right)+k_{\Gamma}\left(q_{i},q_{j}\right),
\]
with $k_{\Gamma}\left(q_{i},q_{j}\right)$ denoting the covariance
function associated to the path loss of the signal. The last equation
follows from the fact the the sum of two kernels, i.e. $\left(\ref{eq:basic CNR model}\right)$,
results in the sum of the respective covariance functions. To determine
$k_{\Gamma}\left(q_{i},q_{j}\right)$, we follow \cite{neal1998bayesian}
to get
\begin{align*}
k_{\Gamma}\left(q_{i},q_{j}\right) & =\mathbb{E}_{q_{b}}\left[\Gamma\left(q_{i};q_{b}\right)\cdot\Gamma\left(q_{j};q_{b}\right)\right]\\
 & =\int_{\mathbb{R}^{2}}g\left(q_{b}\right)\Gamma\left(q_{i};q_{b}\right)\Gamma\left(q_{j};q_{b}\right)dq_{b}.
\end{align*}
Letting $K_{q}=\left[k\left(q,q_{1}\right),k\left(q,q_{2}\right),\ldots,k\left(q,q_{\ell}\right)\right]^{\top}$
and $K$ a matrix with entries defined by $K_{ij}=k\left(q_{i},q_{j}\right)$,
the posterior estimate of the channel at an unknown location $q$
can be found with mean
\[
\bar{\Upsilon}_{dB}\left(q\right)=K_{q}^{\top}\left(K+\sigma_{\rho}^{2}I\right)^{-1}{\bf y},
\]
with ${\bf y}=\left[y_{1},y_{2},\ldots,y_{\ell}\right]^{\top}$ a
vector of $\ell$ measurements, and $I$ the identity matrix of the
approriate size. The variance of the channel estimate is given by
\[
\Sigma\left(q\right)=k\left(q,q\right)-K_{q}{}^{\top}\left(K+\sigma_{\rho}^{2}I\right)^{-1}K_{q}.
\]
An example is shown in Figure \ref{fig:Channel est example}, where a vehicle collects measurements and uses the above method to estimate the unkown communucation channel.

\section{Results}\label{sec:Results}

\begin{figure*}
\begin{centering}
\subfloat[Channel estimate and first optimal trajectory from a single measurement.\label{fig:Channel-estimate 1}]{\centering{}\includegraphics[width=5.5cm]{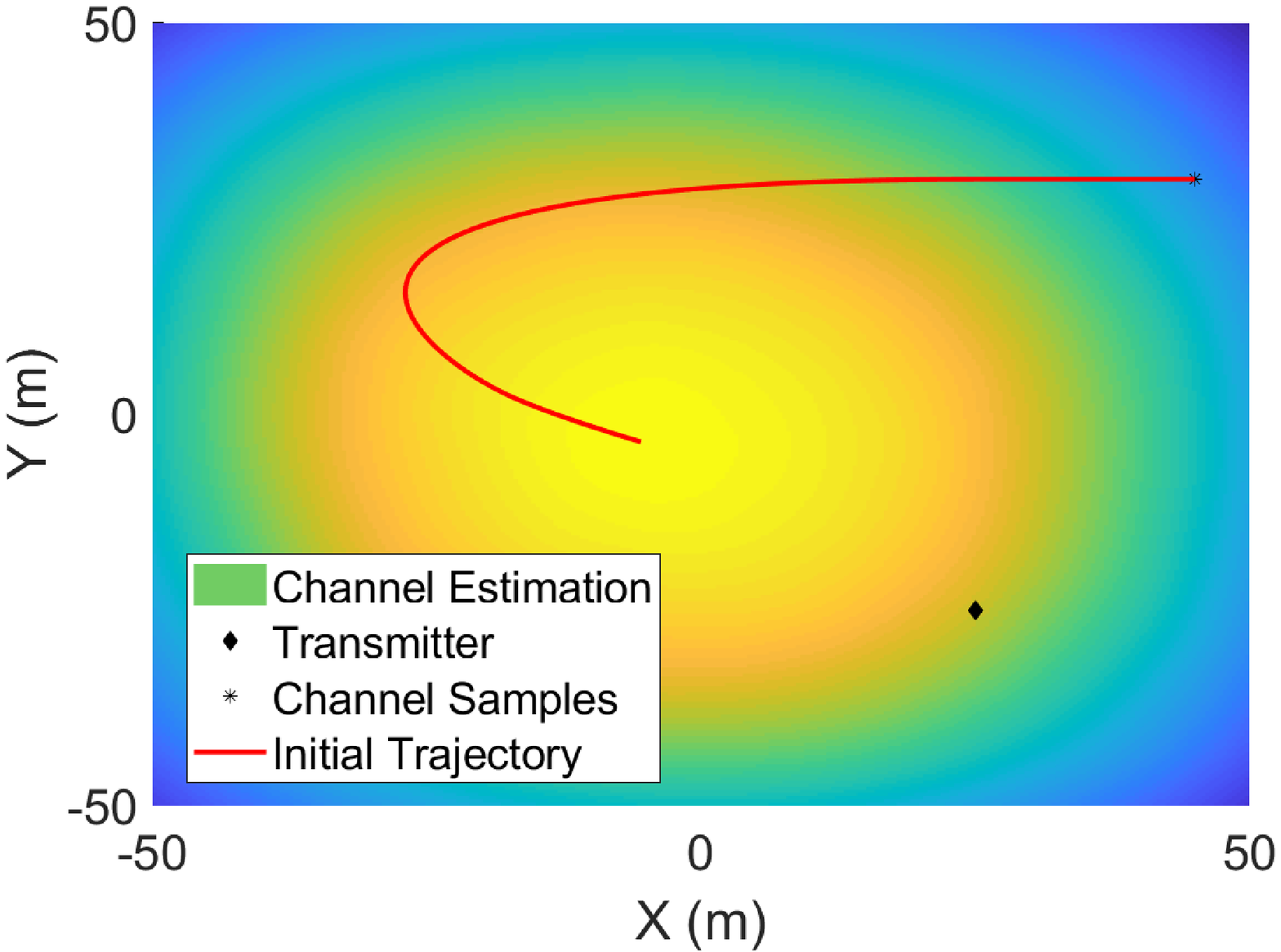}}\hspace*{\fill}\subfloat[After traveling along the first trajectory of $\delta=10$ seconds,
shown in solid black, the vehicle collects the measurements marked
with a '{*}'. The channel estimate is updated and a new optimal trajectory
is computed as shown in green. \label{fig:Channel est 2}]{\begin{centering}
\includegraphics[width=5.5cm]{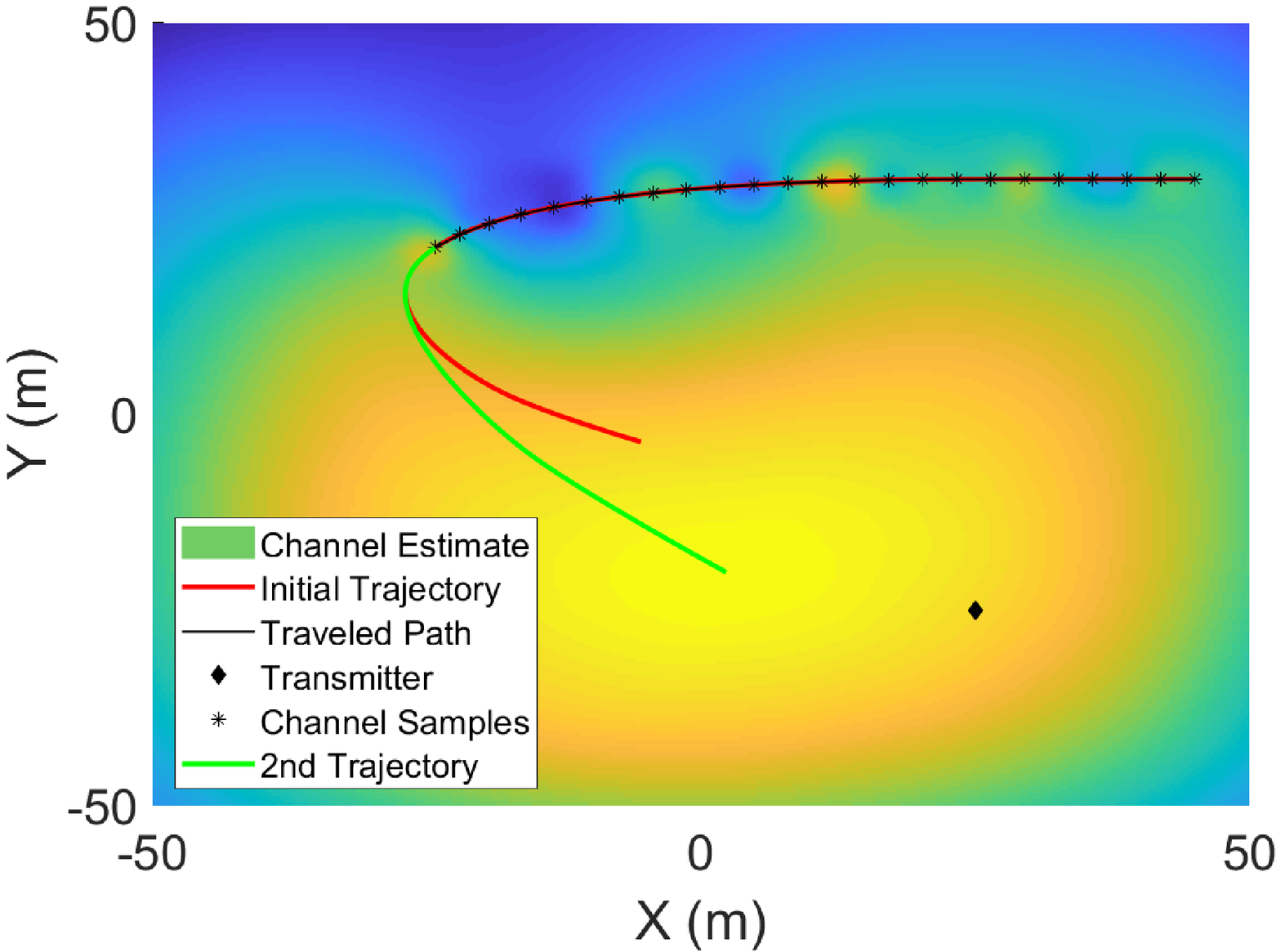}
\par\end{centering}
}\hspace*{\fill}\subfloat[The channel estimate is updated with additional measurements and a
new optimal trajectory is shown in blue. \label{fig:The-channel-estimate 3}]{\centering{}\includegraphics[width=5.5cm]{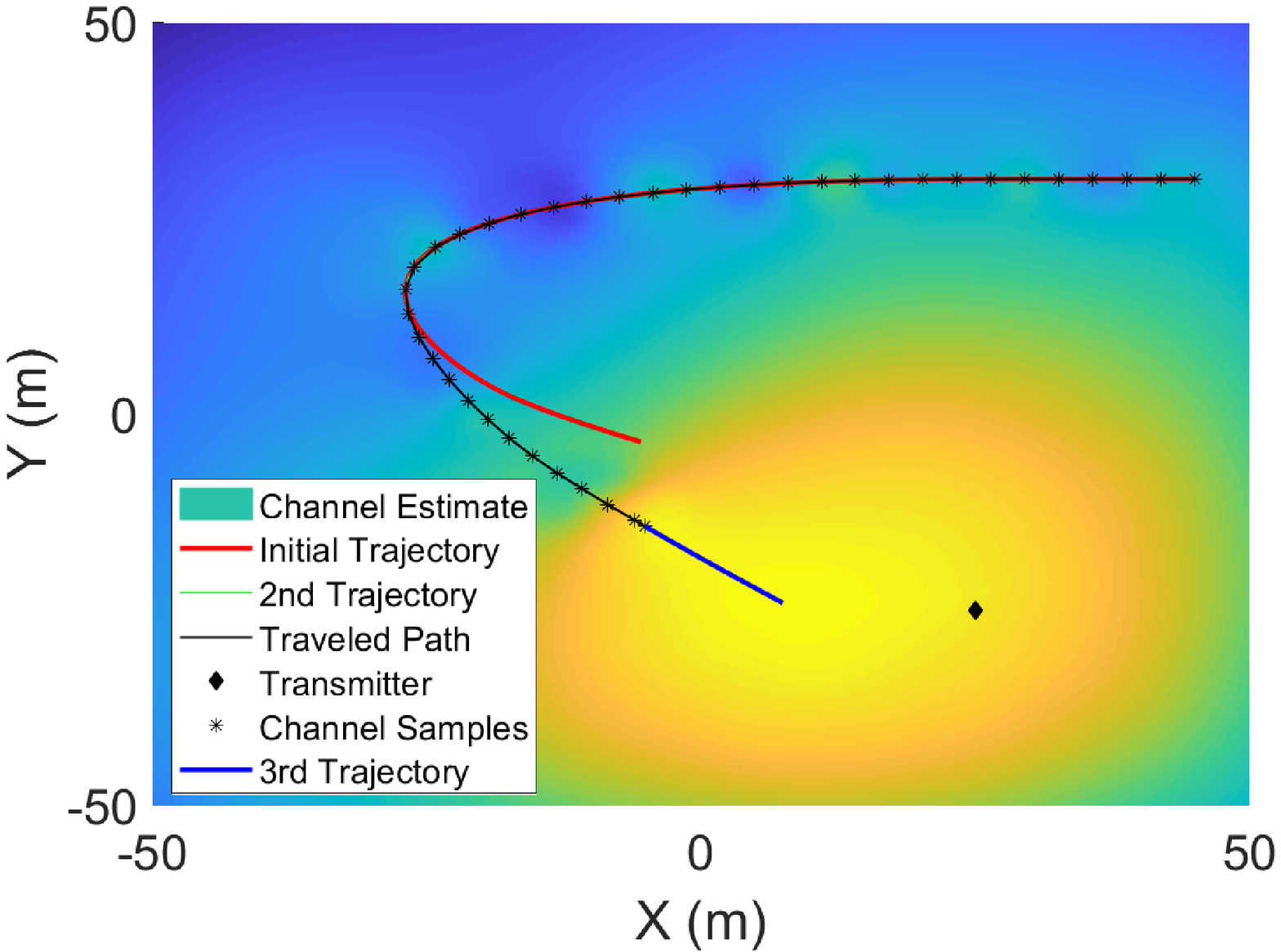}}
\par\end{centering}
\caption{Three optimization cycles of trajectories and their corresponding
channel estimates and optimal trajectories. Best viewed in color.\label{fig:Three-optimization-cycles}}
\end{figure*}

The methods presented above was implemented in MATLAB R2018b on a
laptop equipped with an Intel Core i7-7500 CPU running at 2.70 GHz.
We used as values for the parameters for communication channels what
was presented in \cite{ali2016motion} with $\sigma_{\rho}=1.64$,
$\xi=3.20$ $c_{PL}=-41.34$, $\eta=3.09$, and $n_{PL}=3.86$. While
the time to compute could vary as more measurements are gathered,
we force a fixed value of time delay for consistency. We also follow
\cite{ali2016motion} for these values with $\tau^{k}=2$ seconds
and $\delta^{k}=10$ seconds for all $k$. The newton update of $\left(\ref{eq:Newton iterate}\right)$ and $\left(\ref{eq:Hamiltonian for Newton and delay}\right)$ is stopped when $\varphi(x,t_i) <= 10^{-3}$.

\subsection{Planar Motion Example}

We choose for dynamics $\left(\ref{eq:general linear system}\right)$
with state $x\in\left[q,\dot{q}\right]^{\top}$, where $q\in\mathbb{R}^{2}$
is spatial position of a robot and $\dot{q}\in\mathbb{R}^{2}$ is
the velocity and 
\[
A=\left[\begin{array}{cccc}
0 & 0 & 1 & 0\\
0 & 0 & 0 & 1\\
0 & 0 & 0 & 0\\
0 & 0 & 0 & 0
\end{array}\right],\,B=\left[\begin{array}{cc}
0 & 0\\
0 & 0\\
1 & 0\\
0 & 1
\end{array}\right].
\]
The control $u\in\mathbb{R}^{2}$ is constrained to lie in the set
$\left\Vert u\right\Vert _{2}\leq1$. Since the 2-norm is self-dual,
the Hamiltonian $\left(\ref{eq:Hamiltonian of 'z'}\right)$ for this
example is
\[
\widehat{H}\left(s,p\right)=\begin{cases}
\left\Vert -B^{\top}e^{-sA^{\top}}p\right\Vert _{2} & s\geq\tau\\
-p^{\top}e^{-sA}B\alpha^{k-1}\left(s+\delta^{k-1}\right) & s<\tau
\end{cases},
\]
and the Hamitonian $\left(\ref{eq:Hamiltonian for Newton and delay}\right)$
for the Newton update becomes
\[
H\left(p^{k},x\right)=-x^{\top}A^{\top}p^{k}+Q\left(p^{k}\right),
\]
with 
\[
Q\left(p^{k}\right)=\begin{cases}
\left\Vert -B^{\top}p^{k}\right\Vert _{2} & s\geq\tau\\
-\left(B\alpha^{k-1}\left(s+\delta^{k-1}\right)\right)^{\top}p^{k} & s<\tau
\end{cases}.
\]
The control is found as
\[
\alpha^{k}\left(s\right)=\begin{cases}
\frac{-B^{\top}e^{-sA^{\top}}p^{k}}{\left\Vert -B^{\top}e^{-sA^{\top}}p^{k}\right\Vert _{2}} & s\geq\tau\\
\alpha^{k-1}\left(s+\delta^{k-1}\right) & s<\tau
\end{cases}.
\]
The initial position of the robot is $q_{0}=\left[45,30\right]^{\top}$
with initial velocity $\dot{q}_{0}=\left[-10,0\right]^{\top}$, and
the transmitter is located at $q_{b}=\left[25,-25\right]^{\top}$.
We assume no prior channel measurements before moving, unlike in \cite{ali2016motion},
and use for the prior of transmitter location a uniform distribution
of the operating area, with $q_{b}\sim\mathcal{U}\left(\left[-50,50\right]\times\left[-50,50\right]\right)$.
The vehicle collects an additional sample when the vehicle has
displaced $\eta$ meters, the scale length of $k_{\Delta}$ kernel
function. The vehicle guides to a goal set that is an ellipsoidal
neighborhood around the peak of the estimated CNR by setting
\[
\Omega^{k}=\left\{ x:\left\langle \left(x-\tilde{x}^{k}\right),W^{-1}\left(x-\tilde{x}^{k}\right)\right\rangle \leq1\right\} ,
\]
where $\tilde{x}^{k}=\left(\tilde{q}_{b}^{k},0,0\right)^{\top}$
%\[
%\tilde{x}^{k}=\left[\begin{array}{c}
%\tilde{q}_{b}^{k}\\
%0\\
%0
%\end{array}\right]
%\]
and $\tilde{q}_{b}^{k}$ is the peak of the estimated CNR at the $k$-ith
iteration. The matrix $W$ is positive definite and defines the shape
of the neighborhood. For this example, we use
\[
W=\left[\begin{array}{cccc}
1 & 0 & 0 & 0\\
0 & 1 & 0 & 0\\
0 & 0 & V_{\text{max}}^{2} & 0\\
0 & 0 & 0 & V_{\text{max}}^{2}
\end{array}\right],
\]
with $V_{\text{max}}$ being the maximal allowable velocity at the
goal. We consider a vehicle that comes to close to rest at the goal,
so set $V_{\text{max}}=0.1$. This implies the follow initial value
function
\[
J^{k}\left(x\right)=\left(x-\tilde{x}^{k}\right)^{\top}W^{-1}\left(x-\tilde{x}^{k}\right)-1.
\]
Figure \ref{fig:Three-optimization-cycles} shows the estimation and
associated optimal trajectory for 3 optimization cycles. The vehicle
starts with only a single sample of channel at the vehicle's initial
location. With the transmitter's location unknown and assumed equally
likely at any spot in the operating area, the estimate is biased towards
the center of the area. The estimate and optimal trajectory to the
peak in the estimated channel is shown in Fig. \ref{fig:Channel-estimate 1}.
After traveling along this path for 10 seconds and acquiring more
channel samples, a new estimate is shown in Fig \ref{fig:Channel est 2}.
The path was in an area of low received signal strength and the new
estimate reduces the estimated channel in this region, as well as
shifting the estimate of the peak closer to the ground-truth location.
A time delay of 2 seconds was induced from the channel estimate and
is compensated in the online update to the optimal trajectory, shown
in green. Figure \ref{fig:The-channel-estimate 3} shows another cycle
of more measurements and an improved estimation of the channel with
corresponding optimal path.

\begin{ack}
The author would like to thank Arjun Muralidharan at the University
of California, Santa Barbara (now a software engineer in the Network
Infrastructure Group at Google, Inc.) for the many helpful and enlightening
discussions on communication models used for robotic platforms.
\end{ack}

\bibliography{WP}             % bib file to produce the bibliography

\begin{thebibliography}{24}
\providecommand{\natexlab}[1]{#1}
\providecommand{\url}[1]{\texttt{#1}}
\providecommand{\urlprefix}{URL }
\expandafter\ifx\csname urlstyle\endcsname\relax
  \providecommand{\doi}[1]{doi:\discretionary{}{}{}#1}\else
  \providecommand{\doi}{doi:\discretionary{}{}{}\begingroup
  \urlstyle{rm}\Url}\fi

\bibitem[{Bellman(1957)}]{bellman1957dynamic}
Bellman, R.E. (1957).
\newblock \emph{Dynamic Programming}, volume~1.
\newblock Princeton University Press.

\bibitem[{Bryson and Ho(1975)}]{bryson1975applied}
Bryson, A.R. and Ho, Y.C. (1975).
\newblock \emph{Applied Optimal Control: Optimization, Estimation and Control}.
\newblock CRC Press.

\bibitem[{Camacho and Alba(2013)}]{camacho2013model}
Camacho, E.F. and Alba, C.B. (2013).
\newblock \emph{Model Predictive Control}.
\newblock Springer.

\bibitem[{Darbon and Osher(2016)}]{darbon2016algorithms}
Darbon, J. and Osher, S. (2016).
\newblock Algorithms for overcoming the curse of dimensionality for certain
  {Hamilton}-{Jacobi} equations arising in control theory and elsewhere.
\newblock \emph{Research in the Mathematical Sciences}, 3(1), 19.

\bibitem[{Evans(2010)}]{evans10}
Evans, L.C. (2010).
\newblock \emph{Partial Differential Equations}.
\newblock American Mathematical Society, Providence, R.I.

\bibitem[{Franklin et~al.(2006)Franklin, Powell, and
  Emami-Naeini}]{franklin1994feedback}
Franklin, G.F., Powell, J.D., and Emami-Naeini, A. (2006).
\newblock \emph{Feedback Control of Dynamic Systems}.
\newblock Prentice Hall, 5 edition.

\bibitem[{Hiriart-Urruty and Lemar{\'e}chal(2012)}]{hiriart2012fundamentals}
Hiriart-Urruty, J.B. and Lemar{\'e}chal, C. (2012).
\newblock \emph{Fundamentals of Convex Analysis}.
\newblock Springer Science \& Business Media.

\bibitem[{Hopf(1965)}]{hopf1965generalized}
Hopf, E. (1965).
\newblock Generalized solutions of non-linear equations of first order.
\newblock \emph{Journal of Mathematics and Mechanics}, 14, 951--973.

\bibitem[{Kirchner et~al.(2018{\natexlab{a}})Kirchner, Hewer, Darbon, and
  Osher}]{kirchner2018primaldual}
Kirchner, M.R., Hewer, G., Darbon, J., and Osher, S. (2018{\natexlab{a}}).
\newblock A primal-dual method for optimal control and trajectory generation in
  high-dimensional systems.
\newblock In \emph{IEEE Conference on Control Technology and Applications},
  1575--1582.

\bibitem[{Kirchner et~al.(2018{\natexlab{b}})Kirchner, Mar, Hewer, Darbon,
  Osher, and Chow}]{kirchner2017time}
Kirchner, M.R., Mar, R., Hewer, G., Darbon, J., Osher, S., and Chow, Y.
  (2018{\natexlab{b}}).
\newblock Time-optimal collaborative guidance using the generalized {Hopf}
  formula.
\newblock \emph{IEEE Control Systems Letters}, 2(2), 201--206.

\bibitem[{Kurzhanski and Varaiya(2014)}]{kurzhanski2014dynamics}
Kurzhanski, A.B. and Varaiya, P. (2014).
\newblock \emph{Dynamics and Control of Trajectory Tubes: Theory and
  Computation}, volume~85.
\newblock Springer.

\bibitem[{Kwon and Pearson(1980)}]{kwon1980feedback}
Kwon, W. and Pearson, A. (1980).
\newblock Feedback stabilization of linear systems with delayed control.
\newblock \emph{IEEE Transactions on Automatic control}, 25(2), 266--269.

\bibitem[{Kwon et~al.(2004)Kwon, Lee, and Han}]{kwon2004general}
Kwon, W.H., Lee, Y., and Han, S.H. (2004).
\newblock General receding horizon control for linear time-delay systems.
\newblock \emph{Automatica}, 40(9), 1603--1611.

\bibitem[{Lu(2008)}]{lu2008delay}
Lu, M.C. (2008).
\newblock \emph{Delay Identification and Model Predictive Control of Time
  Delayed Systems}.
\newblock Ph.D. thesis, McGill University.

\bibitem[{Malmirchegini and Mostofi(2012)}]{malmirchegini2010spatial}
Malmirchegini, M. and Mostofi, Y. (2012).
\newblock On the spatial predictability of communication channels.
\newblock \emph{IEEE Transactions on Wireless Communications}, 11(3), 964--978.

\bibitem[{Mitchell(2008)}]{mitchell2008flexible}
Mitchell, I.M. (2008).
\newblock The flexible, extensible and efficient toolbox of level set methods.
\newblock \emph{Journal of Scientific Computing}, 35(2), 300--329.

\bibitem[{Neal(1996)}]{neal1998bayesian}
Neal, R.M. (1996).
\newblock \emph{Bayesian Learning for Neural Networks}, volume 118 of
  \emph{Lecture Notes in Statistics}.
\newblock Springer, New York.

\bibitem[{Osher and Fedkiw(2006)}]{osher2006level}
Osher, S. and Fedkiw, R. (2006).
\newblock \emph{Level Set Methods and Dynamic Implicit Surfaces}, volume 153.
\newblock Springer Science \& Business Media.

\bibitem[{Osmolovskii(1998)}]{osmolovskii1998calculus}
Osmolovskii, N.P. (1998).
\newblock \emph{Calculus of Variations and Optimal Control}, volume 180.
\newblock American Mathematical Society.

\bibitem[{Pontryagin(2018)}]{pontryagin2018mathematical}
Pontryagin, L.S. (2018).
\newblock \emph{Mathematical Theory of Optimal Processes}.
\newblock Routledge.

\bibitem[{Rasmussen(2004)}]{rasmussen2004gaussian}
Rasmussen, C.E. (2004).
\newblock \emph{Gaussian Processes in Machine Learning}.
\newblock Springer.

\bibitem[{Richard(2003)}]{richard2003time}
Richard, J.P. (2003).
\newblock Time-delay systems: an overview of some recent advances and open
  problems.
\newblock \emph{Automatica}, 39(10), 1667--1694.

\bibitem[{Ross et~al.(2006)Ross, Gong, Fahroo, and Kang}]{ross2006practical}
Ross, I.M., Gong, Q., Fahroo, F., and Kang, W. (2006).
\newblock Practical stabilization through real-time optimal control.
\newblock In \emph{American Control Conference, 2006}, 6--pp. IEEE.

\bibitem[{Usman et~al.(2016)Usman, Cai, Mostofi, and Wardi}]{ali2016motion}
Usman, A., Cai, H., Mostofi, Y., and Wardi, Y. (2016).
\newblock Motion and communication co-optimization with path planning and
  online channel prediction.
\newblock In \emph{American Control Conference (ACC), 2016}, 7079--7084. IEEE.

\end{thebibliography}
                                                     % with bibtex (preferred)
                                                   
%\begin{thebibliography}{xx}  % you can also add the bibliography by hand

%\bibitem[Able(1956)]{Abl:56}
%B.C. Able.
%\newblock Nucleic acid content of microscope.
%\newblock \emph{Nature}, 135:\penalty0 7--9, 1956.

%\bibitem[Able et~al.(1954)Able, Tagg, and Rush]{AbTaRu:54}
%B.C. Able, R.A. Tagg, and M.~Rush.
%\newblock Enzyme-catalyzed cellular transanimations.
%\newblock In A.F. Round, editor, \emph{Advances in Enzymology}, volume~2, pages
%  125--247. Academic Press, New York, 3rd edition, 1954.

%\bibitem[Keohane(1958)]{Keo:58}
%\newblock \emph{Power and Interdependence: World Politics in Transitions}.
%\newblock Little, Brown \& Co., Boston, 1958.

%\bibitem[Powers(1985)]{Pow:85}
%\newblock Is there a way out?
%\newblock \emph{Harpers}, pages 35--47, June 1985.

%\bibitem[Soukhanov(1992)]{Heritage:92}
%\newblock \emph{{The American Heritage. Dictionary of the American Language}}.
%\newblock Houghton Mifflin Company, 1992.

%\end{thebibliography}

%\appendix
%\section{A summary of Latin grammar}    % Each appendix must have a short title.
%\section{Some Latin vocabulary}              % Sections and subsections are supported  
                                                                         % in the appendices.
\end{document}